\documentstyle[prb,twocolumn,aps]{revtex}
\begin{document}
\draft
\title{\bf Charge dynamics in the colossal 
magnetoresistance pyrochlore Tl$_2$Mn$_2$O$_7$}
\author{H. Okamura, T. Koretsune, M. Matsunami, 
S. Kimura,\cite{PRESTO} T. Nanba}
\address{Physics Department and Graduate School of 
Science and Technology, Kobe University, 
Kobe 657-8501, Japan}
\author{H. Imai, Y. Shimakawa, Y. Kubo}
\address{Fundamental Research Laboratories, 
NEC Corporation, Tsukuba 305-8501, Japan}

\date{\today}
\maketitle
\begin{abstract}

Optical conductivity data [$\sigma(\omega)$] of the colossal 
magneto-resistance (CMR) pyrochlore Tl$_2$Mn$_2$O$_7$ are 
presented as functions of temperature ($T$) and external 
magnetic field ($B$).   
Upon cooling and upon applying $B$ near the Curie 
temperature, where the CMR manifests itself, 
$\sigma(\omega)$ shows a clear transition from an 
insulatorlike to a metallic electronic structure as 
evidenced by the emergence of a pronounced Drude-like 
component below $\sim$ 0.2~eV.     Analyses on the 
$\sigma(\omega)$ spectra show that both $T$- 
and $B$-induced evolutions of the electronic structure 
are very similar to each other, and that they are 
universally related to the development of 
macroscopic magnetization ($M$).     In particular, 
the effective carrier density obtained from $\sigma(\omega)$ 
scales with $M^2$ over wide ranges of $T$ and $B$.    
The contributions to the CMR from the carrier effective 
mass and scattering time are also evaluated from 
the data.  
\end{abstract}


Physics of colossal magneto-resistance (CMR) phenomena 
has been one of the central issues of condensed matter 
physics in the last decade.     
In particular, the ferromagnetic perovskite 
manganites, e.g., La$_{1-x}$Sr$_x$MnO$_3$, have attracted 
much attention.\cite{tokura}     
In these compounds, a Mn$^{3+}$/Mn$^{4+}$ double 
exchange interaction reduces the transfer energy of 
the Mn 3$d$ holes through a parallel alignment of 
neighboring Mn spins,\cite{zener} resulting in a 
CMR near the Curie temperature ($T_c$).    
In addition, a strong Jahn-Teller effect due to 
Mn$^{3+}$ leads to the formation 
of polarons, which strongly affects the transport 
properties.\cite{mills}    
More recently, the Tl$_2$Mn$_2$O$_7$ pyrochlore has been 
attaining increasing interest, since it exhibits a CMR 
that is comparable to those observed for the 
perovskites.\cite{shimakawa,subram,cheong}    
Tl$_2$Mn$_2$O$_7$ is also a ferromagnet, and its 
resistivity ($\rho$) drops rapidly upon cooling 
below $T_c$~$\sim$ 120~K.   
Near and above $T_c$, an external magnetic field of 
7~T reduces $\rho$ by a factor of $\sim$ 10.  
Although these features appear very similar to those 
for the perovskites, various studies have suggested 
that the underlying mechanism should be very different: 
In Tl$_2$Mn$_2$O$_7$ the electric conduction takes 
place in a conduction band having strong Tl 6$s$ and 
O 2$p$ characters, as shown by band 
calculations.\cite{singh,mishra,shimakawa3}   The spontaneous 
magnetization below $T_c$ is produced by the 
Mn$^{4+}$ sublattice through superexchange interaction, 
independently from the conduction system.    
The CMR results from changes in the conduction system 
caused by the magnetization in the Mn$^{4+}$ sublattice.    
Little evidence has been found for a double exchange or a 
Jahn-Teller effect in Tl$_2$Mn$_2$O$_7$.   
Hall effect experiments\cite{shimakawa,imai} have shown 
that the conduction electron density is very low, 
$n \sim 0.8 \times 10^{19}$~cm$^{-3}$ or 0.001 per formula 
unit (f.u.) above $T_c$, and that it increases to 
$n \sim 5 \times 10^{19}$~cm$^{-3}$ (0.006 per f.u.) 
in the ferromagnetic phase.   A density increase is 
also found with applied magnetic field.\cite{imai}    
The carrier density change has been considered 
a main cause for the CMR in Tl$_2$Mn$_2$O$_7$.\cite{imai,martinez}

However, many questions remain regarding the CMR 
mechanism in Tl$_2$Mn$_2$O$_7$.   First of all, the 
microscopic electronic structure of Tl$_2$Mn$_2$O$_7$ 
itself has remained unclear experimentally, since no 
spectroscopic data have been available, to our 
knowledge, over a sufficiently wide energy range.  
Among many other issues, for example, the contribution 
to the CMR from a change in the carrier mobility, 
suggested by the much larger variation in $\rho$ 
than in $n$,\cite{imai} has not been studied 
in detail.

In this work we use optical spectroscopy to probe the 
microscopic electronic structure of Tl$_2$Mn$_2$O$_7$.    
The optical conductivity $\sigma(\omega)$ 
of Tl$_2$Mn$_2$O$_7$ was obtained from its optical 
reflectivity 
$R(\omega)$, measured at photon energies 0.007~eV 
$\leq \hbar\omega \leq$ 30~eV, at temperatures 40~K 
$\leq T \leq$ 300~K, in magnetic fields $B \leq$ 6~T.   
Upon cooling through $T_c$ and upon applying $B$ 
near $T_c$, $\sigma(\omega)$ indicates a transition 
from insulatorlike to metallic electronic structures 
with a large Drude-like component below 0.2~eV.  
It is found that the $B$-induced evolution of the 
electronic structures leading to the CMR is very 
similar to the $T$-induced one, and that they are 
universally related to the macroscopic magnetization.   
This is the first demonstration of such microscopic electronic 
structures in Tl$_2$Mn$_2$O$_7$ as a function of energy.    
Contributions to the CMR from changes in the carrier 
mobility are also analyzed.

The Tl$_2$Mn$_2$O$_7$ sample used was a 
polycrystalline disk\cite{poly} synthesized by solid state 
reaction under a pressure of 2~GPa.\cite{shimakawa}    
The measured magnetization of the sample showed a 
sharp onset at $T_c$ $\sim$ 120~K, and near and 
above $T_c$ the resistivity decreased by a factor 
of $\sim$ 10 by applying a field of 6~T.   
These values agree well with those in the 
literature.\cite{shimakawa,subram,cheong,shimakawa3,imai}    
The sample surface was mechanically polished for optical 
studies.    Near-normal incidence reflectivity measurements 
were made using a Fourier interferometer and conventional 
sources for $\hbar \omega \leq$ 2.5~eV, and using synchrotron 
radiation source for $\hbar \omega \leq$ 30~eV.    
A gold or a silver film evaporated directly onto the sample 
surface was used as a reference of the reflectivity below 
2.5~eV.    
The experiments under magnetic fields were done using a 
superconducting magnet.       $\sigma(\omega)$ 
spectra were obtained from the measured $R(\omega)$ using 
the Kramers-Kronig relations.\cite{wooten}     
To complete $R(\omega)$, a Hagen-Rubens or a constant 
extrapolation was used for the lower-energy end, and a 
$\omega^{-4}$ extrapolation for the higher-energy 
end.\cite{wooten}

Figure 1(a) shows $R(\omega)$ and $\sigma(\omega)$ 
spectra of Tl$_2$Mn$_2$O$_7$ at 295~K up to 30~eV.    
The spectra are typical of an insulating (semiconducting) 
oxide, with very small $\sigma(\omega)$ below 1~eV and 
sharp peaks below 0.1~eV due to optical 
phonons,\cite{phonon} and indicate a small 
density of states (DOS) around the Fermi level ($E_F$).   
From $\sigma(\omega)$, the band gap of Tl$_2$Mn$_2$O$_7$ at 
295~K is estimated to be $\sim$ 1.6~eV.    
The peaks in the spectra above 2~eV can be attributed to 
charge transfer excitations from O 2$p$ to Mn 3$d$ bands 
(the strong peak near 2.5~eV) and excitations to other 
higher-lying states.\cite{singh,mishra,shimakawa3}     
Figure 1(b) shows $R(\omega)$ and $\sigma(\omega)$ 
below 0.5~eV at several $T$'s.\cite{footnote4}    
Upon cooling below $T_c$=120~K, both $R(\omega)$ and 
$\sigma(\omega)$ increase rapidly, with $R(\omega)$ 
approaching 1 and $\sigma(\omega)$ showing a sharp 
rise toward the lower energy end.   The emergence of 
this pronounced Drude-like component clearly 
demonstrates that the electronic structure 
near $E_F$ becomes metallic below $T_c$, with a 
substantial number of free carriers.   
The spectral weight of a Drude component in $\sigma(\omega)$ 
can be related to the effective density of free carriers, 
$N_{eff}$, through the optical sum rule\cite{wooten} as 
\begin{equation}
N_{eff}=\frac{n}{m^\ast} = 
\frac{2 m_0}{\pi e^2} \int_0^{\omega_p} \sigma_D(\omega) d\omega.  
\end{equation}
Here, $n$ is the carrier density, $m^\ast$ is the 
effective mass in units of the rest electron mass $m_0$, 
$\omega_p$ is the plasma cut-off energy, and 
$\sigma_D(\omega)$ is the Drude contribution in 
$\sigma(\omega)$.   In the inset of Fig.~1(b) we plot 
$N_{eff}$, calculated using (1) as a function of 
$T$.\cite{plasma}   
Contributions from the phonon peaks in $\sigma(\omega)$ 
have been subtracted by fitting them with the Lorentz 
function.\cite{wooten}   The measured magnetization 
($M$) of the same sample is also plotted.    
The increase in $N_{eff}$ is apparently synchronous 
with that in $M$ below $T_c$, 
showing a strong connection between the dynamic charge 
response and $M$.

According to the band calculations for ferromagnetic 
(FM) Tl$_2$Mn$_2$O$_7$,\cite{singh,mishra,shimakawa3} 
there is a conduction band with its bottom located 
$\sim$ 0.5~eV below $E_F$.\cite{shimakawa3}  
This conduction band results from a strong hybridization 
of Tl 6$s$, O 2$p$, and Mn 3$d$ states, and the 
resulting effective mass is small, $m^\ast \simeq m_0$.   
The DOS around $E_F$ is very small, and the calculated 
electron density is 4 $\times 10^{19}$~cm$^{-3}$ (0.005 
per f.u.).   The minimum $k$-conserving gap in the DOS, 
which leads to an optical gap in the dipole 
approximation,\cite{wooten} is 1.5-2~eV in both the majority- 
and minority-spin band structures.  The observed 
$\sigma(\omega)$ below $T_c$ is consistent with these 
predictions: In the observed $R(\omega)$ and 
$\sigma(\omega)$ the metallic components are indeed 
limited to below 0.5~eV.   The onset of 
$\sigma(\omega)$ was observed at $\sim$ 1.6~eV also 
below $T_c$ (not shown), similarly to that above $T_c$ 
shown in Fig.~1.    (The spectra above 0.5~eV showed 
only minor $T$ dependences.)   From the measured 
$N_{eff}$ of 0.01 per f.u. below $T_c$ [see Fig.~1(b)] 
and the measured density of $n$ = 0.006 per 
f.u.,\cite{shimakawa,imai} we obtain 
$m^\ast \sim 0.6m_0$, which is indeed small and close 
to the calculated $m^\ast$.    Hence, the band 
calculations\cite{singh,mishra,shimakawa3} are 
successful in predicting the basic electronic 
structures of FM Tl$_2$Mn$_2$O$_7$ near $E_F$.

Figures 2 and 3 show $R(\omega)$ and $\sigma(\omega)$, 
respectively, in external magnetic fields $B \leq$ 6~T, 
applied normal to the sample surface.\cite{footnote5}    
The spectral changes at 125~K, near $T_c$, are quite 
spectacular: both $R(\omega)$ and $\sigma(\omega)$ 
show large increases with $B$, and $\sigma(\omega)$ shows 
a clear transition from an insulating character at 0~T 
to a metallic one at 6~T.    
In contrast, at 160~K and 40~K, away from $T_c$, they 
show only small changes.    The $B$-induced spectral 
changes at 125 K are remarkably similar to those 
observed with decreasing $T$ at $B$=0 shown in Fig.~1(b).   
This similarity, emphasized in the bottom graph of Fig.~3, 
clearly demonstrates that the electronic structures near 
and above $T_c$ under strong $B$ fields are very similar 
to those below $T_c$ at $B$=0.    
Figure~4(a) shows the variation of $N_{eff}$, calculated 
with (1), as a function of $B$ relative to that at $B$=0.     
The relative increase in $N_{eff}$ is largest in 
the range 120~K $\leq T \leq$ 140~K, 
which is exactly where the measured CMR of the 
sample was most pronounced.

In Fig.~4(b) we plot $N_{eff}$ as a function of $M$ for 
different values of $T$ and $B$.    Apparently, 
{\it $N_{eff}$ has a universal relation with $M$} 
in quite wide ranges of $T$ and $B$.    
The solid curve in Fig.~4(b) is a quadratic fit to 
the data, which demonstrates that {\it $N_{eff}$ 
nearly scales with $M^2$}.    
Note that this single universal relation 
holds in two seemingly distinct regimes, i.e., 
one below $T_c$ where $\rho$ depends more strongly 
on $T$ than on $B$, and the other near and above $T_c$ 
where $\rho$ depends more strongly on $B$ than on $T$.  
This common universal relation, as well as the similarity 
between $T$-induced and $B$-induced spectral changes, 
indicate that the electronic structures and the 
dynamic charge response in the presence of a macroscopic 
$M$ are basically common in the two regimes.      
Namely, near and above $T_c$ (120 - 140~K), an 
applied external field produces $M$ by aligning the 
Mn$^{4+}$ spins, 
which results in polarized band structures and the 
appearance of a small conduction band, similarly to 
those below $T_c$ predicted by the band 
calculations.\cite{singh,mishra,shimakawa3}    
Here, as our results indicate, $M$ is the essential 
parameter for the microscopic electronic structures 
around $E_F$, while $T$ and $B$ are not.    
The CMR in Tl$_2$Mn$_2$O$_7$ near $T_c$ should be 
primarily due to this {\it appearance of conduction 
band in the presence of $M$ induced by external $B$}.   
The significance of the observed $N_{eff} \propto M^2$ 
scaling relation will be discussed later.  

Regarding the CMR of Tl$_2$Mn$_2$O$_7$, as stated 
before, contributions from changes in $m^\ast$ and 
the average scattering time ($\tau$) of the carriers 
have been unclear.    Below we will roughly estimate 
these contributions using our data.    
The measured $n$ increases upon cooling through $T_c$, 
$n(100~K)/n(140~K) \sim 5$.\cite{imai,footnote}  
For $N_{eff} = n/m^\ast$, the corresponding ratio is 
$\sim$ 8 [see Fig.~1(b)].     These values suggest 
that $m^\ast$ becomes $8/5 \sim$ 1.5 times smaller 
below $T_c$.     The increase in the dc conductivity 
($\sigma_{dc}$) of our sample was 
$\sigma_{dc}(100~K)/\sigma_{dc}(140~K) \sim 20$.   
In the Drude theory, 
$\sigma_{dc} \propto n\tau/m^\ast$.     Comparison 
with $N_{eff}=n/m^\ast$ shows that $\tau$ becomes 
$20/8 \sim$ 2.5 times longer below $T_c$.     
At 140~K on going from $B$=0 to 6~T, $n$ increases 
by a factor of 2,\cite{imai} while $N_{eff}$ and 
$\sigma_{dc}$ increase by factors of 5 and 10, 
respectively.   A comparison of these factors show 
that $\tau$ becomes 2 times longer and $m^\ast$ becomes 
2.5 times smaller at 6~T.    
Hence, {\it $\tau$ becomes greater and $m^\ast$ 
becomes smaller both by a factor of $\sim$ 2}, both 
in the $T$-induced and external $B$-induced FM phases.  
Namely, all of $n$, $\tau$, and $m^\ast$ contribute to 
the CMR in Tl$_2$Mn$_2$O$_7$.     
Note that the above simple 
analyses\cite{footnote2} are possible since the 
Drude-like component in $\sigma(\omega)$ is completely 
separated from those due to interband transitions.

Since $n$ is very small above $T_c$, 
$\sim 0.8 \times 10^{19}$~cm$^{-3}$,\cite{imai} 
a metallic conduction with a well-defined band 
effective mass is unlikely.    
Indeed, experimental evidence for a hopping conduction 
has been reported,\cite{martinez} and a possibility of 
magnetic polarons has been discussed.\cite{littlewood}    
Hence the observed decrease of ``optical'' $m^\ast$ 
might be related to a crossover from a hopping-dominated 
charge dynamics to a Drude-like one.    The increase 
in $\tau$ is likely to result mainly from the magnetic 
critical scattering of carriers by localized 
spins,\cite{langer} where the strong spin fluctuations 
present for $T \geq T_c$ are suppressed for $T < T_c$, 
leading to larger $\tau$ and $\sigma_{dc}$.   
Effects of a local coupling 
between conduction electrons and Mn spins above $T_c$ 
have also been discussed.\cite{imai}

As stated earlier, our $\sigma(\omega)$ data of 
Tl$_2$Mn$_2$O$_7$ have exhibited an $N_{eff} \propto M^2$ 
scaling relation for the Drude-like component in the FM phase.     
A microscopic origin of this $M^2$ scaling is unclear 
currently.    Note that a 
relative shift of rigid, spin-up and spin-down parabolic 
bands would lead to an $M^{1.5}$ scaling,\cite{imai} 
rather than $M^2$.   
Hence, the appearance of $M^2$-scaled spectral weight 
transfer in Tl$_2$Mn$_2$O$_7$ is non-trivial, and should 
be understood based on the specific microscopic electronic 
structures of Tl$_2$Mn$_2$O$_7$.     
An analogous $M^2$-scaled spectral weight transfer 
has been also reported for other magnetoresistance (MR) 
ferromagnets, the perovskite manganites\cite{moritomo-sugai} 
and EuB$_6$.\cite{degiorgi}     
In the case of the perovskites, the $M^2$-scaled spectral 
weight transfer was consistently explained by a theory 
based on the double exchange.\cite{furukawa}    
However, since the double exchange is unimportant in 
Tl$_2$Mn$_2$O$_7$, the mechanism of the $M^2$ scaling 
for Tl$_2$Mn$_2$O$_7$ is likely to be distinct from that 
for the perovskites.     Nevertheless, it is 
remarkable that these ferromagnets, with different 
mechanisms for their large MR effects, all show 
an $M^2$-scaled spectral weight transfer.     
It remains an open, interesting question whether this 
scaling is a universal optical property of MR ferromagnets 
regardless of the detailed origin for their ferromagnetism 
and large MR effects.

In summary, our optical data for the CMR pyrochlore 
Tl$_2$Mn$_2$O$_7$ have revealed much new information 
about its microscopic electronic structures as functions 
of energy, $T$, and $B$.      
With increasing $B$ and with decreasing $T$ near $T_c$, 
$\sigma(\omega)$ has shown a dramatic transition from an 
insulatorlike to a metallic electronic structure, with the 
appearance of a small ($\sim$ 0.2~eV from $E_F$) conduction 
band.    
The $\sigma(\omega)$ data show that the $B$-induced evolution 
of the electronic structure leading to the CMR is 
very similar to the $T$-induced one, and that they are 
universally related to $M$ in wide ranges of $T$ and $B$.   
In particular, $N_{eff}$ scales with $M^2$.   
Our data have also shown that $\tau$ becomes greater 
and $m^\ast$ becomes smaller both by a factor of $\sim$ 2 
in the FM phase, contributing to the CMR.

We would like to thank L. Degiorgi for a preprint on 
his work on EuB$_6$.    
The experiments using synchrotron radiation source 
was done at the UVSOR Facility, Institute for Molecular 
Science.    We thank the staff for assistance.

\begin{figure}
\caption{(a) Optical reflectivity ($R$) and conductivity 
($\sigma$) of Tl$_2$Mn$_2$O$_7$ at 295~K.  (b) $R$ and 
$\sigma$ measured at different temperatures at zero field.  
The inset compares the effective carrier density ($N_{eff}$) 
and the magnetization ($M$) at an external magnetic field 
of 0.2~T, as a function of temperature ($T$). }
\end{figure}

\begin{figure}
\caption{Optical reflectivity ($R$) of Tl$_2$Mn$_2$O$_7$ 
in magnetic fields at several temperatures.  }
\end{figure} 

\begin{figure}
\caption{ Top four graphs: optical conductivity ($\sigma$) 
of Tl$_2$Mn$_2$O$_7$ in magnetic fields at several 
temperatures.   Bottom graph: $\sigma$ at different 
temperatures and magnetic fields plotted together, 
demonstrating their similarity (see the text).    }
\end{figure} 

\begin{figure}
\caption{(a) Effective carrier density ($N_{eff}$) as 
a function of magnetic field ($B$), normalized by 
those at $B$=0.    (b) $N_{eff}$ at different 
temperatures and magnetic fields as a function of 
magnetization ($M$).   The solid curve is a quadratic 
fit to the data.     The vertical bars indicate 
$N_{eff}$ arising from the range below 
0.007~eV, where the reflectivity spectra were 
extrapolated.  }
\end{figure} 


\begin{references}
\bibitem{poly} 
The crystal structure of pyrochlore Tl$_2$Mn$_2$O$_7$ 
is cubic.   Hence the electronic structures are expected 
to be isotropic, and the use of a polycrystalline 
sample should not be a problem.
\bibitem{footnote4} $R(\omega)$ above 2.5~eV was measured 
at 295~K only, to which $R(\omega)$ below 2.5~eV at 
lower $T$'s were smoothly connected.
\bibitem{plasma} We obtained $\omega_p$ for each 
$\sigma(\omega)$ as $\omega$ satisfying both 
$\epsilon_1(\omega)$=0 and 
$\frac{d\epsilon_1(\omega)}{d\omega}>$0, where 
the real dielectric function $\epsilon_1(\omega)$ 
was calculated from $R(\omega)$.\cite{wooten}
\bibitem{footnote5} $R(\omega)$ in $B$ fields were 
measured below 0.9~eV only, where $R(\omega)$ 
showed only minor $B$-dependence above $\sim$ 0.5~eV.
\bibitem{footnote} The quality of our sample is 
very similar to that used in Ref.~\onlinecite{imai}, 
and we have assumed that the carrier densities in 
the two works are equal.
\bibitem{footnote2} 
Alternatively, we could also estimate $\tau$ and $m^\ast$ 
as adjustable parameters in fitting $\sigma(\omega)$ 
to a Drude function.    In this treatment the tail of 
a Drude component contains important information, but 
this range also contains strong phonon peaks in our data.    
Consequently we have found that the procedure based on (1) 
is more plausible than the above one.    
Full details of the analysis will be presented elsewhere.
\end{references}
\end{document}